\documentclass[conference]{IEEEtran}
\IEEEoverridecommandlockouts
\usepackage{amsmath,amssymb,amsfonts}
\usepackage{graphicx}
\usepackage[fontsize=11pt]{fontsize}
\usepackage[obeyFinal]{todonotes}
\usepackage{booktabs}
\usepackage{microtype} %

\newtheorem{definition}{Definition}[section]

\usepackage{xspace} %

\newcommand\vsm[1]{\text{$\textit{v}_{\text{sm},#1}$}\xspace} %
\def\vsmn{\text{$\textit{v}_{\text{sm},\textit{i}}$}\xspace} %

\def\vout{\text{$\textit{v}_{\text{out}}$}\xspace} %
\def\iout{\text{$\textit{i}_{\text{out}}$}\xspace} %

\def\fiout{\text{$f_{i_{\text{out}}}$}\xspace} %
\def\fvout{\text{$f_{v_{\text{out}}}$}\xspace} %

\def\Sa{\text{$\text{S}_{1}$}\xspace} %
\def\Sb{\text{$\text{S}_{2}$}\xspace} %
\def\Sc{\text{$\text{S}_{3}$}\xspace} %
\def\Sd{\text{$\text{S}_{4}$}\xspace} %

\newcommand{\ins}{\mathsf{insertion}}
\newcommand{\forw}{\mathsf{forward}}
\newcommand{\back}{\mathsf{backward}}
\newcommand{\byOne}{\mathsf{bypass 1}}
\newcommand{\byTwo}{\mathsf{bypass 2}}
\newcommand{\by}{\mathsf{bypass}}

\newcommand{\B}{\mathsf{B}}
\newcommand{\F}{\mathsf{I}}

\newcommand{\vpdot}[1]{\dot{\textit{v}}_{\text{p,\(#1\)}}} %
\newcommand{\vp}[1][k]{\text{${\textit{v}}_{\text{p,\(#1\)}}$}\xspace} %
\newcommand{\Rp}[1][k]{\text{${\textit{R}}_{\text{p,\(#1\)}}$}\xspace} %
\newcommand{\oRp}[1][k]{\text{$\overline{{\textit{R}}}_{\text{p,\(#1\)}}$}\xspace} %
\newcommand{\Cp}[1][k]{\text{${\textit{C}}_{\text{p,\(#1\)}}$}\xspace} %
\newcommand{\oCp}[1][k]{\text{$\overline{{\textit{C}}}_{\text{p,\(#1\)}}$}\xspace} %
\newcommand{\Ro}[1][k]{\text{${\textit{R}}_{\text{o,\(#1\)}}$}\xspace} %
\newcommand{\oRo}[1][k]{\text{$\overline{{\textit{R}}}_{\text{o,\(#1\)}}$}\xspace} %

\newcommand{\Em}[1][k]{\text{$\textit{v}_{\text{ocv,\(#1\)}}$}\xspace} %
\newcommand{\oEm}[1][k]{\text{$\overline{\textit{v}}_{\text{ocv,\(#1\)}}$}\xspace} %

\newcommand{\vcell}[1]{\text{$\textit{v}_{\text{cell,\(#1\)}}$}\xspace} %
\newcommand{\icell}[1]{\text{$\textit{i}_{\text{cell,\(#1\)}}$}\xspace} %

\newcommand{\eqn}[2]{\text{\(e_{#1, #2}\)}} %
\newcommand{\fcell}[1]{\text{${f}_{\text{cell,\(#1\)}}$}\xspace} %

\newcommand{\fltcell}[1]{\text{\(f_{\text{cell},#1}\)}\xspace} %
\newcommand{\flt}[1]{\text{\(f_{#1}\)}\xspace} %
\newcommand{\sens}[1]{\text{\(y_{#1}\)}\xspace} %

\usepackage{caption}
\usepackage{subcaption}
\usepackage{graphicx} 
\usepackage{prettyref} %
\newrefformat{fig}{Fig.~\ref{#1}}
\newrefformat{tab}{Table~\ref{#1}}

\newlength{\bibitemsep}
\newlength{\bibparskip}
\let\oldthebibliography\thebibliography
\renewcommand\thebibliography[1]{%
	\oldthebibliography{#1}%
	\setlength{\parskip}{\bibitemsep}%
	\setlength{\itemsep}{\bibparskip}%
}
\graphicspath{{figures/}}

\begin{document}

\title{
	Structural Diagnosability Analysis of Switched and Modular Battery Packs
\thanks{*This work is financed by the Swedish Electromobility Center and the Swedish Energy Agency.}
}

\author{\IEEEauthorblockN{1\textsuperscript{st} Fatemeh Hashemniya}
	\IEEEauthorblockA{\textit{Department of Electrical Engineering} \\
		\textit{Linköping University}, Sweden \\
		e-mail: fatemeh.hashemniya@liu.se}
	\and
	\IEEEauthorblockN{2\textsuperscript{nd} Arvind Balachandran}
	\IEEEauthorblockA{\textit{Department of Electrical Engineering} \\
		\textit{Linköping University}, Sweden \\
		e-mail: arvind.balachandran@liu.se}
	\and \IEEEauthorblockN{3\textsuperscript{rd} Erik Frisk}
	\IEEEauthorblockA{\textit{Department of Electrical Engineering} \\
		\textit{Linköping University}, Sweden \\
		e-mail: erik.frisk@liu.se}
	\and
	\IEEEauthorblockN{4\textsuperscript{th} Mattias Krysander}
	\IEEEauthorblockA{\textit{Department of Electrical Engineering} \\
		\textit{Linköping University}, Sweden \\
		e-mail: mattias.krysander@liu.se}
	}
\maketitle
\vspace*{-8mm}
\begin{abstract}
    
    Safety, reliability, and durability are targets of all engineering systems, including Li-ion batteries in electric vehicles. 
    This paper focuses on sensor setup exploration for a battery-integrated modular multilevel converter (BI-MMC) that can be part of a solution to sustainable electrification of vehicles.
    BI-MMC contains switches to convert DC to AC to drive an electric machine. 
    The various configurations of switches result in different operation modes, which in turn, pose great challenges for diagnostics.
    The study explores diverse sensor arrangements and system configurations for detecting and isolating faults in modular battery packs. Configurations involving a minimum of two modules integrated into the pack are essential to successfully isolate all faults.
    The findings indicate that the default sensor setup is insufficient for achieving complete fault isolability. Additionally, the investigation also demonstrates that current sensors in the submodules do not contribute significantly to fault isolability. Further, the results on switch positions show that the system configuration has a significant impact on fault isolability. A combination of appropriate sensor data and system configuration is important in achieving optimal diagnosability, which is a paramount objective in ensuring system safety.
    
\end{abstract}    %


\section{Introduction}
\label{Intro}
Switched systems are commonly found in many applications, e.g., power electronics, automotive, and aerospace. This work will explore how to perform diagnosability analysis in a switched battery system for electric vehicles.

Electric vehicle (EV) powertrains employ two-level inverters to convert DC from a large battery pack to AC that powers an electric machine \cite{poorfakhraei2021review}.
The large battery pack typically contains several series and parallel connected low-voltage battery cells (2-4\,V) to provide high-voltage (300-800\,V) \cite{goli2021review}.  
Due to differences in leakage currents and cell in-homogeneities, individual cell voltage and state-of-charge (SOC) distribution among the cells are non-homogeneous \cite{berg2015batteries}. 
As a result, over time, some cells tend to discharge faster than other cells, thus limiting the total energy the pack can deliver. 
One approach to mitigate this problem is to introduce modular battery packs, a technique that has gained popularity in the research and development of EV powertrains due to their high efficiency and providing cell-level control \cite{quraan2015design, balachandranlic2022}.  
Modular battery packs can improve battery balancing and provide better battery fault tolerance due to their highly modular structure \cite{8515660}. 
However, the increased flexibility makes fault monitoring more demanding than monitoring traditional battery packs because modular battery packs include more components, leading to more types of faults, and can be operated in diverse operation modes which makes diagnostics more challenging. 

The purpose of fault diagnosis is to detect and mitigate faults, i.e., small deviations from nominal behavior, such that failures can be avoided by taking proper actions \cite{blanke2016diagnosis}.
Cell faults include unexpected rates of aging such as increased internal resistance or capacity fade and increased connector resistance. 
Internal/external short circuits and thermal runaways in cells are failures since these imply a permanent interruption of the battery's ability to perform and thus are not the focus of this work. 
Hu et al. \cite{9205673} comprehensively reviewed mechanisms, features, and diagnoses of various faults in Li-ion batteries. %
In addition, battery management systems (BMS) include a wide range of sensors that are essential for battery monitoring and control \cite{plett2015battery}.
These sensors include voltage, current, and temperature sensors and the detection and isolation of sensor faults are also important for the diagnostics \cite{6862432}. Offset and scaling faults are common faults in voltage and current sensors \cite{zheng2020fault}. Therefore, establishing a proper sensor setup to monitor battery conditions is essential.

\section{Problem formulation}
\label{sec:prob-form}

The main problem studied in this paper is how to do fault diagnosis analysis for switched systems and apply it to a modular battery pack model and, in particular, discuss questions concerning computational complexity.
Model-based techniques based on structural analysis \cite{blanke2016diagnosis,frisk2017toolbox} is a suitable choice of methods for addressing this problem, however, extensions to standard analysis techniques are required to include also switched models.

The first contribution of this paper is a method to determine the influence of the total number and type of sensors required to detect and isolate faults in a modular battery pack, utilizing methods for switched systems. 
The diagnostic performance is given not only by the operational mode of an individual submodule (SM) but also by the configuration of all SMs in the battery system. Therefore, a method for investigating how fault detectability and isolability depend on switch configuration and sensor setup is demonstrated.
A second contribution is two techniques for the reduction of the combinatorial complexity in the analysis introduced by the switches. 
A closely related work is \cite{Schmid2021Structural} where two sensor setups have been analyzed, among other things, to reach the best isolability for a reconfigurable battery system. The approach is based on
an equivalent circuit model (ECM) with a half-bridge converter and a two-state thermal model.
One difference to \cite{Schmid2021Structural} is that here the effects of switching between system configurations, leading to different model structures and isolability possibilities, are analyzed.

\section{Diagnosis Background} \label{Background}

To perform diagnostics using structural analysis, a structural model \cite{blanke2016diagnosis} in the form of an incidence matrix is introduced based on a mathematical model of the system. The rows in the incidence matrix represent the set $M$ of system equations and the columns represent the unknown variables. The presence of a variable in an equation is denoted by '1', and its absence is denoted by '0'. 
The key analysis tool, the Dulmage-Mendelsohn (DM) decomposition~\cite{dulmage1958coverings}, partitions the incidence matrix into three parts, an underdetermined part, $M^-$, where the number of equations is less than the number of variables, a just-determined part, $M^0$, where the number of equations and variables are equal, and an overdetermined part, $M^+$, where the number of equations is more than the number of variables. The overdetermined part contains the analytical redundancy needed for diagnosis. %

\begin{figure}[t!]
	\centering
	\includegraphics[width=0.48\textwidth]{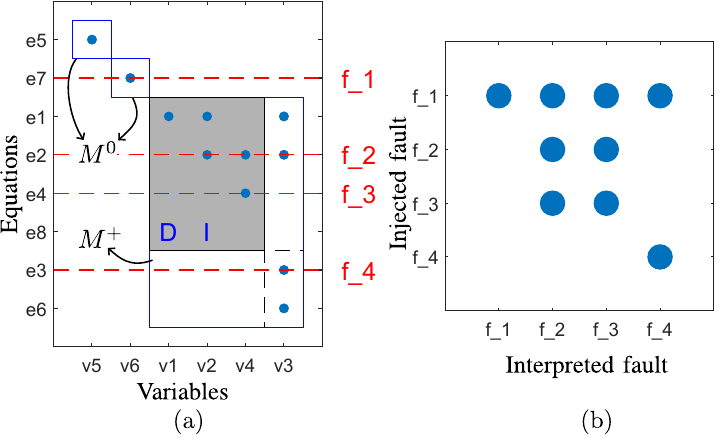}
	\caption{(a) Extended Dulmage–Mendelsohn decomposition and (b) the corresponding isolability matrix.}
	\label{fig:ISO_DM_EX}
\end{figure}

Without loss of generality, assume that each fault only affects one equation and let $e_f$ denote the equation that is affected by a fault $f$. Then the structural fault detectability and isolability are defined as %
\begin{definition}[Structural detectability \cite{Krysander2008sensor}]\label{def:Structural detectable}
	A fault $f$ is structurally detectable in model $M$ if $e_f \in M^+$. \hfill $ \blacklozenge $
\end{definition}
Fig.~\ref{fig:ISO_DM_EX}(a) shows an example of an extended Dulmage-Mendelsohn decomposition \cite{Krysander2008sensor} of an incidence matrix. The just-determined part consists of the two first rows and columns, and the remaining rows and columns are the overdetermined part, i.e., $M^+ = \{e_1, e_2, e_3, e_4, e_6, e_8\}$. The faults $f_2 \text{ to } f_4$  are structurally detectable since they are in $M^+$ but not $f_1$. The next definition characterizes structural model properties required for being able to point out which fault has occurred.
\begin{definition}[Structural isolability \cite{Krysander2008sensor}]\label{def:Structural_isolable}
	A fault $f_i$ is structurally isolable from $f_j$ in a model $M$ if $e_{f_i} \in (M \backslash \{e_{f_j}\})^+$. \hfill $ \blacklozenge $ 
\end{definition}
A fault is uniquely isolable in a model if it is structurally isolable from all other faults in a model. A model has full isolability if all faults in the model are uniquely isolable. 
Structural isolability can be efficiently determined using the Dulmage-Mendelsohn decomposition \cite{dulmage1958coverings}, efficiently implemented in, e.g., the \texttt{dmperm} command in MATLAB and also available in \cite{frisk2017toolbox}.
The isolability of a model can be visualized by partitioning the equations in $M^+$ such that faults are isolable if and only if they are in different equation sets in the partition~\cite{Krysander2008sensor}. In Fig.~\ref{fig:ISO_DM_EX}(a), $M^+ = \{e_1, e_2, e_4, e_8\}$ $\cup$ $\{e_3\}$ $\cup$ $\{e_6\}$ where gray boxes indicate sets with cardinality greater than one, in this case, there is only one such set. Thus, the figure shows that $f_4$ is isolable from all other faults, i.e., $f_4$ is uniquely isolable, and $f_2$ and $f_3$ are isolable from $f_4$ but not isolable from each other. 

Fault isolability can condensly be represented with an isolability matrix. Fig.~\ref{fig:ISO_DM_EX} (b) shows the isolability matrix corresponding to the model presented in Fig.~\ref{fig:ISO_DM_EX}(a). 
A dot in position $(f_i, f_j)$ implies that $f_i$ is not isolable from $f_j$ and full isolability corresponds to an identity matrix.

\section{Modular Battery Pack: An Overview} \label{Topology}
\begin{figure}[t!]
	\centering
	\includegraphics[width=0.48\textwidth]{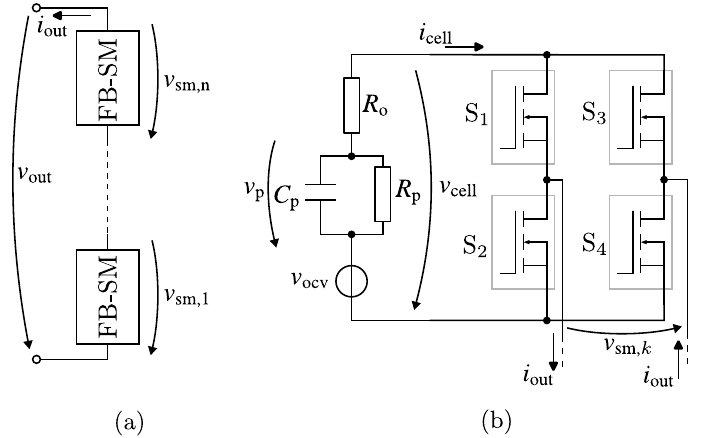}
	\caption{Battery-integrated modular multilevel converter (BI-MMC) phase-leg with $n$ full-bridge submodules. (a) BI-MMC phase-arm and (b) full-bridge submodule.}
	\label{fig:BIMMC_FBSM}
\end{figure}

The battery pack considered works as a battery-integrated modular multilevel converter (BI-MMC) with several cascaded stages of DC-AC converters called submodules (SMs). 
A BI-MMC SM has a few series- and parallel-connected cells defined by the output voltage (\vout) and the total energy stored in the battery pack \cite{balachandranlic2022}. 
The SMs are controlled in such a way that a sinusoidal output voltage is achieved that is used to drive an electric machine \cite{balachandranlic2022}. 
To control the speed and torque of the electric machine, the output current (\iout) is required \cite{harnefors2014control} and this is measured using a current sensor. 
\prettyref{fig:BIMMC_FBSM}(a) shows the schematic of a BI-MMC phase-leg with $n$ full-bridge (FB) SMs and the schematic of the FB-SM is shown in \prettyref{fig:BIMMC_FBSM}(b). 
The cells in the SM are modeled as a lumped RC equivalent circuit model with one RC link. 
This model can capture the degradation of Li-ion batteries \cite{olofsson2014impedance}.
Cell models with constant phase elements can be used to effectively capture and distinguish the various degradation phenomena at the cost of increased complexity and computational time \cite{skoog2017parameterization}.
To estimate the SOC, crucial for any BMS, information about the cell current (\icell{k}) and voltage (\vcell{k}) are necessary \cite{plett2015battery}, where $k$ corresponds to the $k$:th SM. 
The \vcell{k} is measured using a voltage sensor and \icell{k} is either measured, using a current sensor, or estimated, using \iout.

\begin{figure}[b!]
	\centering
	\includegraphics[width=0.40\textwidth]{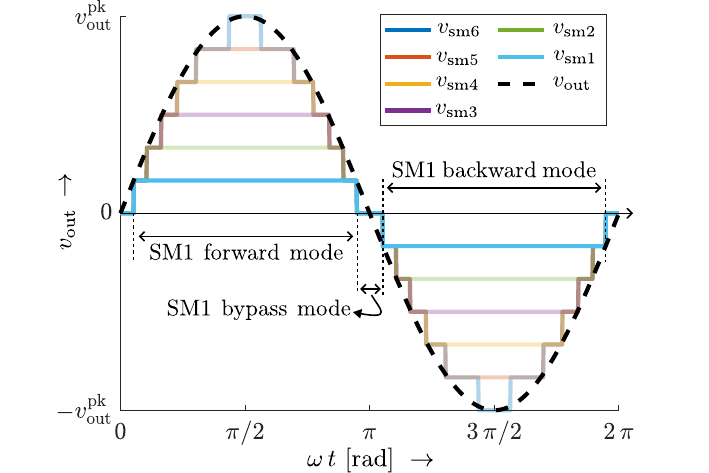}
	\caption{Nearest level modulation for a BI-MMC phase arm with 6 full-bridge submodules.}
	\label{fig:NLCM}
\end{figure} %

The FB-SM, presented in \prettyref{fig:BIMMC_FBSM}(b) has two complimentary switch pairs, \Sa \Sb, and \Sc \Sd, %
and by controlling the `on'- and `off'-durations of these switch pairs, called modulation, an AC output voltage can be achieved. 
Typically, in MMCs (also applicable to BI-MMCs) with a large number of SMs, nearest-level modulation is chosen because of its low switching frequency, easy implementation, and low total harmonic distortion \cite{kouro2007high}. 
\prettyref{fig:NLCM} shows an illustration of the nearest level modulation of a BI-MMC phase arm with 6 FB-SMs.
From the figure, the SM operation can be classified into three modes, based on the output voltage and the states of the SM switches. 
The different modes of the FB-SM based on the `on' and `off' states of the switches are summarized in \prettyref{tab:SMswmodes}. 
\begin{table}[t!]
    \centering
    \caption{Full-bridge submodule modes.}
    \label{tab:SMswmodes}
    \begin{tabular}{l|c|c|c|c|c}
        mode&\Sa & \Sb & \Sc & \Sd & \vsm{k}\\
        \hline
        $ \forw $ &on  & off & off & on  & \vcell{k}\\
        $ \byOne $&on  & off & on  & off & 0 \\
        $ \back $ &off & on  & on  & off  & $-$\vcell{k}\\
        $ \byTwo $&off & on  & off & on  & 0\\
    \end{tabular}
\end{table} 
\section{Modelling for Fault Diagnosis}
\label{sec:model}
This section presents the cell and BI-MMC models for four different sensor setups and their corresponding fault equations.
Each SM is modeled as a set of differential-algebraic equations, and the total number of equations depends on the sensor setups.
The nomenclature, \eqn{x}{k} refers to equation number \(x\) in the \(k\):th SM. Faults are modeled as fault signals $f_i$ where $f_i \neq 0$ if the fault is present and $f_i = 0$ otherwise.

It is important to note that the model presented in this section is a highly idealized and simplistic model of battery cell behavior \cite{skoog2017parameterization}. However, since the analysis is based on structural properties, the method is directly applicable to more detailed models.

\subsection{Model with sensor setup I}
Sensor setup I has voltage sensors in every SM to measure \vcell{k}, where $k=1, \dots, n$ representing $n$ number of SMs, and a current sensor measuring \iout.
In this sensor setup, there are 10 equations per SM.
The cells are modeled by
\begin{equation*}
	\begin{aligned}
		&	\eqn{1}{k}: &\vpdot{k} &= \frac{\icell{k}}{\Cp} - \frac{\vp}{\Rp\,\Cp},\\
		&	\eqn{2}{k}: &\vcell{k} &= \vp + \Ro\,\icell{k} + \Em,\ \text{and}\\
		&	\eqn{3}{k}: &\vpdot{k} &= \frac{d\vp}{dt},
	\end{aligned}
\end{equation*}
where \vp is the voltage across the double-layer capacitance, \Rp is the charge transfer resistance, \Cp is the double-layer capacitance, \Ro is the ohmic resistance, and \Em is the open circuit voltage of the cell in the \(k\):th SM.
These parameters depend on the cells' SOC, state of health (SOH), and temperature, which are here assumed known entities.
During the assumed diagnosis period, a couple of output fundamental AC periods, the parameters can with good accuracy be assumed constant.
The SOC, SOH, and temperature dependency in the parameters, \Ro, \Rp, \Cp, and \Em are, e.g., modeled using look-up tables.  %

The equations describing faults in the behavior of the cells are
\begin{equation*}
	\begin{aligned}
		&	\eqn{4}{k}: &\Ro &= \oRo + \flt{\Ro},\\
		&	\eqn{5}{k}: &\Cp &= \oCp + \flt{\Cp},\\
    	&	\eqn{6}{k}: &\Rp &= \oRp + \flt{\Rp}, \text{and}\\
		&	\eqn{7}{k}: &\Em &= \oEm + \flt{\Em},
	\end{aligned}
\end{equation*}
where \oRo, \oCp, \oRp, and \oEm are the values of the corresponding nominal parameters for present state of charge and temperature, and the corresponding fault signals $\flt{\Ro}$, $\flt{\Cp}$, $\flt{\Rp}$, and, $\flt{\Em}$ models a deviation from the corresponding nominal value.
If $|f_c| \geq \mathrm{L}$, where $c$ is a component and $\mathrm{L}$ is a predetermined limit, e.g., state of health limit, then $c$ is considered faulty. %
The four inner cell faults change gradually, representing aging phenomena \cite{skoog2017parameterization}. A general cell fault $\fltcell{k}$ will be considered instead of considering the internal faults $\{\flt{\Ro}, \flt{\Cp}, \flt{\Rp}, \flt{\Em}\}$ of the cell individually. 
The voltage sensor in every SM to measure \vcell{k} is modeled by
\begin{equation*}
	\begin{aligned}
		&	\eqn{8}{k}: &\sens{\vcell{k}} &= \vcell{k} + \flt{\vcell{k}},
	\end{aligned}
\end{equation*}
where $\sens{\vcell{k}}$ is the sensor signal, $\vcell{k}$ the measured voltage, and $\flt{\vcell{k}}$ a possible sensor fault.

The different modes of the SM, $m_k\in \{\forw,\back, \byOne, \byTwo\}$, make the BI-MMC a multi-mode system. The mode-dependent equations are,
\begin{equation*}\label{eq:switches}
	\begin{aligned}
		&   \eqn{9}{k}:    & \vsm{k} &= \begin{cases}
			\vcell{k},     & \text{if } m_k \in \{ \forw\}\\
			-\vcell{k},    & \text{if } m_k \in \{ \back\}\\
			0,          & \text{o.w.,}
		\end{cases}\\
		&	\eqn{10}{k}:   &\icell{k}     &=  \begin{cases}
			\iout,      & \text{if } m_k \in \{ \forw\}\\
			-\iout,     & \text{if } m_k \in \{ \back\}\\
			0,          & \text{o.w.,}
		\end{cases}
	\end{aligned}
\end{equation*}
where \vsm{k} is the output voltage of the $k$:th SM.

The equations describing the BI-MMC output voltage $\vout$ and current \iout are,
\begin{equation*}\label{eq:global}
	\begin{aligned}
		&	\eqn{1}{0}: &\vout &= \sum_{k=1}^{n} \vsm{k},
		&  &\eqn{2}{0}: &y_{\iout} &= \iout + \flt{\iout},\\
	\end{aligned}
\end{equation*}
where $y_\iout$ is the current measurement, and $f_\iout$ models the current sensor fault.
For $n$ number of SMs, there are $10 n$ equations (\eqn{\{1, \dots, 10\}}{k}, where $k = 1, \dots, n$) describing the faults and operation of SMs, and two more (\eqn{\{1, 2\}}{0}) describing the \vout and \iout, i.e., this sensor setup has $10 n + 2$ equations and the fault set is $\{\fltcell{k},\flt{\vcell{k}}\}_{k=1}^n \cup \{\flt{\iout}\}$.

\subsection{Model with sensor setup II}
Sensor setup II, in addition to the sensors used in setup I, has a sensor measuring the output voltage, i.e., $\vout$. %
In this sensor setup, there are 10 equations per SM %
$\{\eqn{i}{k}\}_{i=1}^{10}$, the output voltage and current equations $\{\eqn{i}{0}\}_{i = 1}^2$, and the sensor equation  
\begin{equation*}\label{eq:global_scI}
	\begin{aligned}
		& \eqn{3}{0}: & y_{\vout} &= \vout + f_{\vout},
	\end{aligned}
\end{equation*}
where $y_\vout$ is the voltage measurement and $f_\vout$ models a voltage sensor fault.
This setup has $10n + 3$ equations, and the fault set is $\{\fltcell{k},\flt{\vcell{k}}\}_{k=1}^n \cup \{\flt{\iout},\flt{\vout}\}$.

\subsection{Model with sensor setup III}
Sensor setup III has voltage and current sensors in each SM, i.e., $\{\vcell{k}, \icell{k}\}_{k = 1}^n$ are measured. The output current \iout is also measured. This sensor setup has 2 output equations $\{\eqn{i}{0}\}_{i = 1}^2$, and 11 equations per SM, i.e., $\{\eqn{i}{k}\}_{i = 1}^{11}$ where 
\begin{equation*}\label{eq:sm_scII}
	\begin{aligned}
		&	\eqn{11}{k}: &y_{\icell{k}} &= \icell{k} + \flt{\icell{k}},\\
	\end{aligned}
\end{equation*}
is a current sensor equation, $y_{\icell{k}}$ the current measurement, and $\flt{\icell{k}}$ the current sensor fault. This setup has $11 n + 2$ equations and the fault set is $\{\fltcell{k},\flt{\vcell{k}},\flt{\icell{k}}\}_{k=1}^n \cup \{\flt{\iout}\}$.

\subsection{Model with setup IV}
Sensor setup IV includes all both current and voltage sensors on each SM and output current and voltage sensors, i.e., the set of measured variables is $\{\icell{k}, \vcell{k}\}_{k = 1}^n \cup \{\iout, \vout\}$. This sensor setup has 11 equations per SM $\{\eqn{i}{k}\}_{i = 1}^{11}$ and 3 output equations $\{\eqn{i}{0}\}_{i=1}^3$, thus bringing the total number of equations to $11 n + 3$. The fault set for this sensor setup is $\{\fltcell{k}, f_{\vcell{k}}, f_{\icell{k}}\}_{k=1}^n \cup \{f_{\iout}, f_{\vout}\}$.

\section{Isolability analysis method} \label{method}

This section introduces techniques to reduce the combinatorial complexity introduced by the switches.
First, configuration reduction to reduce computational complexity is discussed followed by a method on how to compute the fault detectability and isolability properties of a BI-MMC for different sensor setups and reduced system configurations. 
An efficient representation of the fault isolability properties of the modular battery pack is also presented.

\subsection{SM modes and system configurations reduction}\label{sec:configurations}

In the structural analysis of the BI-MMC, two possible reduction techniques are explored to reduce computational complexity: SM mode reduction and system configuration reduction. Table~\ref{tab:SMswmodes} shows the four modes for each SM and then in a system with $n$ submodules, there are $4^n$ different system switch configurations.

Different switch configurations have different diagnosis properties, the analysis must then be performed for all different switch configurations and
since the number of system configurations, and therefore also the analysis complexity, grows exponentially with the number of submodules it is important to reduce the number of system modes to be able to scale the analysis.

First, consider the SM mode reduction. Note that the modes $\{\forw, \back \}$ and $\{\byOne, \byTwo \}$ have identical structures, as seen in~$e_{9,k}$ and~$e_{10,k}$. %
The only difference in the model equations in $\forw$ and $\back$ mode is a sign change, i.e., $\vsmn = \pm\vcell{k}$, which does not change the structure. Also, the bypass modes have the same structure.
Therefore, both the modes $\{\forw, \back \}$ will hereafter be represented with $\ins$ ($\F$) and $\{\byOne, \byTwo \}$ with $\by$ ($\B$).

Second, consider the system configuration reduction. The number of system configurations to analyze can be reduced since isolability properties are determined only by the number of SMs in $\F \text{ or } \B$ mode and not by the specific modes of SMs. Consider for example $n = 3$ where system configurations are represented as a triple $(m_1,m_2,m_3)\in\{\F, \B\}^3$ where $m_k$ denotes the mode of the $k$:th SM. Then the configurations  $\F\F\B$, $\F\B\F$, and $\B\F\F$ have similar fault isolability properties. This reduces the number of system configurations to be analyzed in a system with $n$ SMs to $n+1$ system configurations representing $0, 1, \ldots, n$ inserted cells, respectively.

\subsection{Compact isolability representation}
\label{sec:compact-representation}
Assume $S$ is the set of the 4 considered sensor setups described in Section~\ref{sec:model}. For a given sensor setup $s \in S$, there are $n+1$ different system configurations to be analyzed, i.e., for $k\in \{0, 1, \ldots, n\}$ inserted cells. 
Let the structural model for sensor setup $s$ and $k$ inserted cells be denoted $M_k(s)$. Similarly, let the detectable faults and the fault isolability for the case with $k$ inserted cells be denoted by $D_{s,k}$, $I_{s,k}$, respectively. These sets are computed with the fault diagnosis toolbox in~\cite{frisk2017toolbox} as 
\begin{align*}
	D_{s, k} &= \mathtt{Detectability}(M_k(s))\\
	I_{s, k} &= \mathtt{Isolability}(M_k(s)),
\end{align*}
and can be represented as shown in Table~\ref{tab:isolability-results} where the subscripts $s$ and $k$ correspond to rows and columns, respectively. 

The diagnosability, i.e., detectability and isolability, of the BI-MMC is computed by going through all sensor setups and system configurations in two nested loops. 
\begin{table*}[htb]
	\centering
	\caption{Fault isolability with different sensor setups and system configurations.}
	\vspace{-2mm}
	\label{tab:isolability without T sensor}
	\label{tab:isolability-results}
	{\small
		\begin{tabular}{l@{\hspace{1\tabcolsep}}ll|c@{\hspace{0.5\tabcolsep}}cc@{\hspace{0.5\tabcolsep}}cc@{\hspace{0.5\tabcolsep}}c}
			\hline
			&
				\multicolumn{2}{c}{$ \text{Sensor} \downarrow \text{Conf.} \rightarrow $}  & \multicolumn{2}{c}{0 inserted cells} & \multicolumn{2}{c}{1 inserted cell} & 
				\multicolumn{2}{c}{$>1$ inserted cell}\\
			\cmidrule(r){4-5} \cmidrule(r){6-7} \cmidrule(r){8-9} 
			Setup & SM  & pack  & 
			non-D & non-I, $\B$ & 
			non-I, $\B$ & non-I, $\F$ &
			non-I, $\B$ & non-I, $\F$ \\
			\hline
			\hline
			I &
			$\vcell{k}$  &$\iout$ &%
			$\{\flt{\iout}\}$ & %
			$\{\fcell{k}, \flt{\vcell{k}}\}$ &%
			$\{\fcell{k}, \flt{\vcell{k}}\}$ & %
			$\{ \fcell{k}, f_{\vcell{k}}, f_{\iout}\}$  & %
			$\{ \fcell{k}, \flt{\vcell{k}}\}$ & %
			$\{ \fcell{k}, \flt{\vcell{k}}\}$ \\ %
			II &
			$\vcell{k}$  & $\iout, \vout$ & %
			$\{\flt{\iout}\}$& %
			$\{ \fcell{k}, \flt{\vcell{k}}\}$ & %
			$\{\fcell{k}, \flt{\vcell{k}}\}$ & %
			$\{\fcell{k}, f_{\iout}\}$ & %
			$\{\fcell{k}, \flt{\vcell{k}}\}$ & %
			$\emptyset$\\
			III &
			$\vcell{k}, \icell{k}$ &$\iout$ & %
			$\{\flt{\iout}\}$& %
			$\{\fcell{k}, \flt{\vcell{k}}\}$ & %
			$\{\fcell{k}, \flt{\vcell{k}}\}$ & %
			$\{\fcell{k}, \flt{\vcell{k}}\}$ & %
			$\{\fcell{k}, \flt{\vcell{k}}\}$ & %
			$\{\fcell{k}, \flt{\vcell{k}}\}$ \\ %
			IV &
			$\vcell{k}, \icell{k}$ &$\iout, \vout$ & %
			$\{\flt{\iout}\}$ & %
			$\{\fcell{k}, \flt{\vcell{k}}\}$ & %
			$\{\fcell{k},\flt{\vcell{k}}\}$ & %
			$\emptyset$ & %
			$\{\fcell{k}, \flt{\vcell{k}}\}$ & %
			$\emptyset$ \\
			\hline
	\end{tabular}}
\end{table*}
Each row Table~\ref{tab:isolability-results} corresponds to a sensor setup that is specified by the first three columns. Sensors that are included in all SMs are listed in the SM column, and sensors measuring the output of the battery pack are listed in the pack column. %

The column labeled non-D under 0 inserted cells specifies the non-detectable faults for each sensor setup, i.e., $\fiout$ is not detectable for any sensor setup with 0 inserted cells. All faults are detectable for all sensor setups if at least one cell is inserted. 

The fault isolability for a model (without taking causality \cite{frisk2012diagnosability} into account) can be specified with a partition of the detectable faults where each fault set in the partition indicates the faults that are not isolable from each other. Thus, a compact representation of the isolability used here is to list the sets of non-isolable faults. 

\begin{figure}
				\centering
				\includegraphics[trim=0 0 0 5, clip, width=0.40\textwidth]{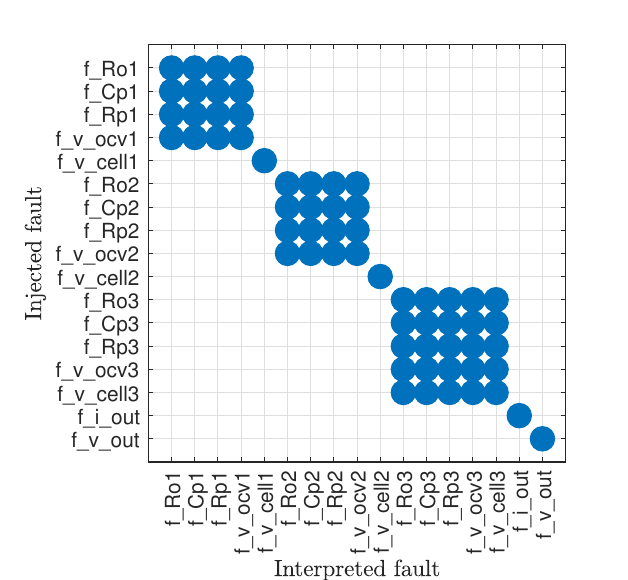}
				\caption{Isolability matrix for BI-MMC with 3SMs in $ \F\F\B $ system configuration with sensor setup II.}
				\label{fig:DM_with_i_cell}
			\end{figure}

To show an example of the isolability of a system and how it is compactly represented in the table, let us consider a switched battery pack with $n = 3$ SMs in mode $\F\F\B$, i.e., with two inserted cells and with sensor setup II, i.e., with voltage sensors in the SMs and voltage and current output sensors. The extended DM decomposition and the corresponding fault isolability matrix are shown in Fig.~\ref{fig. DM plus v pack} and Fig.~\ref{fig:DM_with_i_cell} respectively. Here, the different faults $\{\flt{\Ro}, \flt{\Cp}, \flt{\Rp}, \flt{\Em} \}$ related to the cell parameters are used to fulfill the assumption that faults only affect one equation in the structural analysis. However, when analyzing the result the general cell fault $\fltcell{k}$, introduced in Section~\ref{sec:model}, is considered to represent any type of faulty behavior of the cell. The non-isolable fault sets correspond to the diagonal blocks in the fault isolability matrix, i.e., 
\begin{equation}\label{eq:iso_example}
	\begin{aligned}	I_{\text{II}, 2} = \{ &\overbrace{\{ \fltcell{1}\}, \{\flt{\vcell{1}}\}}^{\text{SM 1}}, 
		\overbrace{\{ \fltcell{2}\}, \{\flt{\vcell{2}}\}}^{\text{SM 2}},\\ 
		& \underbrace{\{ \fltcell{3}, \flt{\vcell{3}}\}}_{\text{SM 3}}, \{\flt{\iout}\}, \{\flt{\vout}\} \}. 
	\end{aligned}
\end{equation}

This example shows some isolability properties that hold for any number of SMs in the system, for any of the considered sensor setups, and for any switch configuration, that can be utilized to compactly describe the isolability in all these different cases. Faults in different SMs are isolable from each other implying that the non-isolable fault sets include faults from at most one SM. Futhermore, SMs in insertion mode, here SM 1 and 2, have similar isolability. SMs in bypass mode have also similar isolability even though it is not shown in this example. Hence, the isolability can be specified by listing the non-isolable fault sets with cardinality strictly greater than 1 for an SM in insertion mode, given in the columns labeled non-I, $\F$ in Table~\ref{tab:isolability-results}, and for an SM in bypass mode given in the columns labeled non-I, $\B$. This implies that faults not present in any non-isolable set are uniquely isolable and full isolability is given by no non-isolable fault sets. 

For the example in~\eqref{eq:iso_example}, there are no non-isolable faults in insertion mode and $\{ \fltcell{3}, \flt{\vcell{3}}\}$ is the non-isolable fault set in bypass mode. This result can be seen in Table~\ref{tab:isolability-results} for setup II and $>$ 1 inserted cell.

\section{Fault isolability result and discussion} \label{isolability analysis}
\label{Section:Fault isolability analysis}
This section will analyze the isolability results summarized in Table~\ref{tab:isolability-results} of modular battery systems for the different sensor setups and conclude with a discussion of how the structural results relate to analytical properties.
	\subsection{Diagnosability for sensor setup I} \label{sec:setup1-result}
		
	    The fault detectability and isolability for sensor setup~I are given in the first row in Table~\ref{tab:isolability-results}, and it can be seen that all faults are detectable in all configurations, except for the case when all SMs are in bypass mode (i.e., 0 inserted cells) where the output current sensor fault $\fiout{}$ is non-detectable. 
		If one cell is inserted, then $\fiout{}$ is detectable but is not isolable from the faults in the inserted cell. If at least two cells are inserted, $\fiout{}$ is uniquely isolable. 
	   	The maximum isolability is achieved when the number of inserted cells is equal to or greater than 2. Then, a fault can be isolated to a specific SM or identified as an output current sensor fault, but the faults within an SM are not isolable from each other.

	\subsection{Diagnosability for sensor setup II}

\begin{figure}[ht]
			\centering
			\includegraphics[trim=0 5 0 22, clip, width=0.48\textwidth]{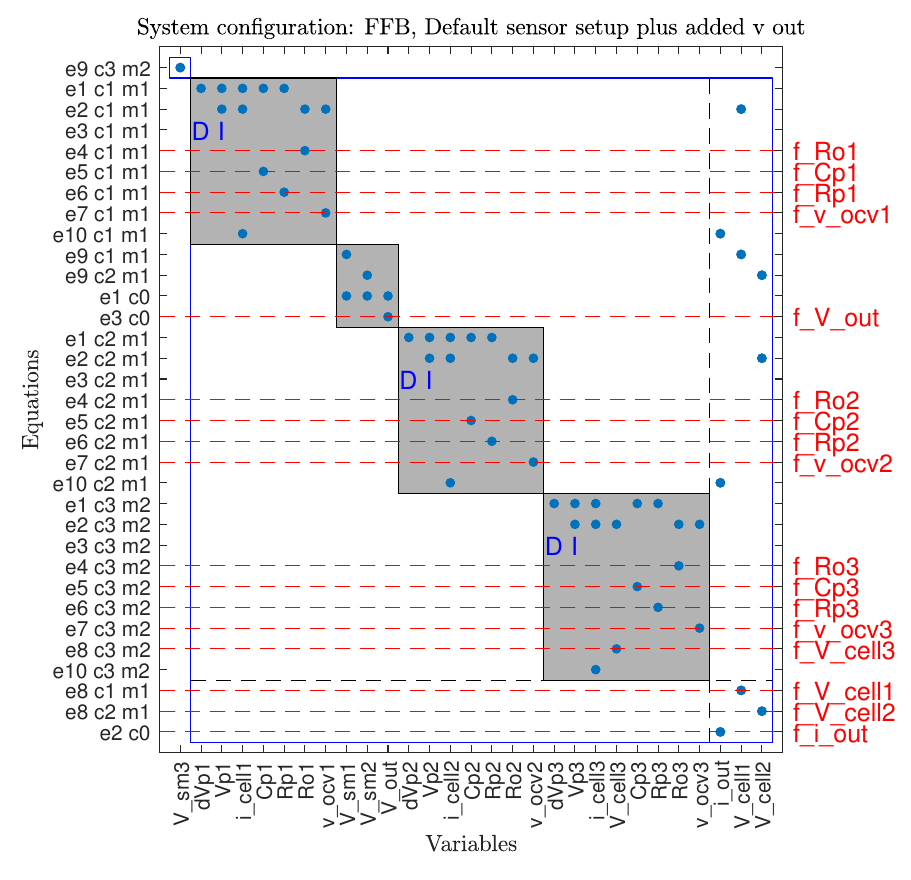}
			\caption{Extended DM decomposition for BI-MMC with 3 SMs in $\F\F\B$ configuration with sensor setup II.}
			\label{fig. DM plus v pack}
		\end{figure}

	The fault detectability and isolability result of sensor setup II
	can be seen on the second row of Table~\ref{tab:isolability-results}. 
	The difference between setup I and II is an output voltage sensor with fault $\fvout$. The output voltage sensor fault $\fvout$ is detectable and uniquely isolable in all system configurations, thus not visible in the table. By adding the output voltage sensor, the faults $\fcell{k}$ and $f_{\vcell{k}}$ are now isolable from each other if and only if SM $k$ is in insertion mode. A more detailed illustration of this property can be seen in Fig.~\ref{fig. DM plus v pack}, which shows the extended DM decomposition for the sensor setup II with the system operated in configuration $\F\F\B$. The faults $f_{\vcell{1}}$ and $f_{\vcell{2}}$ in the SMs in insertion mode are uniquely isolable but not $f_{\vcell{3}}$ in SM 3, which is in bypass mode.

		\subsection{Diagnosability for sensor setup III and IV}
			The isolability results of sensor setup III and IV are given in the third and fourth row in Table~\ref{tab:isolability-results} respectively. 
			The sensor faults in added current sensors $f_{\icell{k}}$ for $k \in \{1, \ldots, n\}$, are detectable and uniquely isolable in both cases. A pairwise comparison of rows 1 with 3 and 2 with 4 shows that the only gain in isolability by adding a current sensor to each SM is that the output current sensor fault $\fiout$ becomes uniquely isolable in the case of one inserted cell. By comparing rows 2 and 3 it is clear that except for the possibility of isolating an output current sensor fault in the case of one inserted cell, it is better to include one output voltage sensor because that will enable the possibility of isolating the cell and voltage sensor faults within each SM in inserted mode.    			

			\subsection{Structural vs analytical isolability}

			Even though the structural detectability or isolability does not improve when sensors are added it can happen that faults can more easily be detected and isolated due to the numerical properties of the model as the next example will show focusing on the detectability of the output current sensor fault $f_{\iout}$. Consider an example system with only one SM where the nominal parameters are 
			$\bar{R}_p = 692$ $\mu\Omega$, $\bar{C}_p = 1.52$ F, $\bar{R}_o = 1.2$ m$\Omega$, and $\bar{\textit{v}}_\text{ocv} = 4.07$ V.   
			For sensor setup I, the following residual ($r$) sensitive to $f_{\iout}$ is valid in the forward and backward mode:
		\begin{equation}\label{eq:res-setup1}
			\begin{aligned}
		\dot{\textit{v}}_p &= \pm \frac{y_{i_\text{out}}}{\bar{C}_p} - \frac{\textit{v}_p}{\bar{R}_p \bar{C}_p},\\
		r &= y_{\textit{v}_\text{cell}} - \textit{v}_p \mp \bar{R}_o y_{i_\text{out}} - \bar{\textit{v}}_\text{ocv}.
		\end{aligned}
		\end{equation}		
		The upper sign is used in forward mode and the one below is in backward mode. In fault-free operation, the residual is 0, but in case of a fault $f_{\iout}$ the residual value can be expressed as a function of the fault signal as 
		\begin{equation}
			\begin{aligned}
				\dot{\textit{v}}_p &= \pm \frac{f_{\iout}}{\bar{C}_p} - \frac{\textit{v}_p}{\bar{R}_p \bar{C}_p}\\
				r &= - \textit{v}_p \mp \bar{R}_o f_{\iout}. 
			\end{aligned}
		\end{equation}		
		For a stationary fault, the gain from fault to residual is $\mp \bar{R}_o$ which is a small number making detection difficult with noisy measurements and model inaccuracies. In bypass mode, the fault $f_\iout$ is not detectable at all as shown previously in %
		Table \ref{tab:isolability without T sensor}.

		Now, consider sensor setup III with an additional sensor $y_{i_\text{cell}}$. Then in addition to residual~\eqref{eq:res-setup1} the following residual can be used for detecting $f_{\iout}$ 
		    \begin{equation}\label{res:icell}
		    	r = y_\iout \mp y_{i_\text{cell}} = f_{\iout} \mp f_{\icell{}}.\\
		    \end{equation}
			The coefficient of $f_{\iout}$ is 1, i.e., the fault sensitivity to $f_{\iout}$ is much better using this residual compared to residual~\eqref{eq:res-setup1}. This is an example of quantitative improvements that are not captured in structural analysis. 
	    	
			Although the addition of current sensors to each cell improves the fault sensitivity of $f_\iout$, there are less expensive solutions to reach the same result. For example, by adding a redundant sensor measuring, the output current
			 \begin{equation}\label{extra_i_sensor}
			 	y^\text{extra}_\iout = \iout + f^\text{extra}_\iout,
			 \end{equation} 
			gives an even better result, since the corresponding residual is
			\begin{equation}\label{res:2ipack}
				r = y_\iout-y^\text{extra}_\iout = \fiout - f^\text{extra}_\iout.\\
			\end{equation}
			The fault sensitivity is the same as in~\eqref{res:icell} but this residual is valid in all system configurations, not only in the insertion modes.

\section{Conclusion} \label{Conclusion}

Investigation on sensor setup and system configuration for BI-MMC for diagnosis purposes was addressed in this paper. A main challenge with this is the combinatorial complexity due to a large number of system switch states. A general method for evaluating fault detectability and isolability in different system configurations was demonstrated. In systems with more than one SM it is sufficient to use voltage sensors in all SMs and an output current and voltage sensor to make all faults uniquely isolable. Full isolability is obtained if all SMs are in insertion mode. The investigation showed that SM current sensors do not significantly improve the structural diagnosability but an additional redundant output current sensor can be used to improve fault sensitivity.

\bibliographystyle{plain}
{\small\bibliography{references.bib}}
\end{document}